\documentclass[10pt,pra,superscriptaddress,twocolumn,longbibliography]{revtex4-2}

\usepackage[T1]{fontenc}
\usepackage{amsmath}
\usepackage{amssymb}
\usepackage{graphicx}
\usepackage{datetime}
\usepackage{xcolor}
\usepackage{newtxtext,newtxmath}

\usepackage{microtype}
\usepackage{comment}
\usepackage{braket}
\usepackage{hyperref}
\hypersetup{%
    bookmarksopen=false,
    bookmarksnumbered=true,
    pdfnewwindow=true,
    unicode=false,
    colorlinks=true,%
    citecolor=blue,
    linkcolor=black,
    urlcolor=blue,
    filecolor=blue
    }

\newcommand{\me}{\mathrm{e}}
\newcommand{\mi}{\mathrm{i}}
\newcommand{\dif}{\mathrm{d}}

\begin{document}

\title{Optimally driving multi-photon transitions in the perturbative single-mode regime} 

\newcommand{\freiburg}{Physikalisches Institut, Albert-Ludwigs-Universit\"{a}t Freiburg, Hermann-Herder-Stra{\ss}e 3, D-79104, Freiburg, Germany}
\newcommand{\zurich}{Institute for Theoretical Physics, ETH Zürich, Zürich 8093, Switzerland}
\newcommand{\qc}{Quantum Center, ETH Zürich, Zürich 8093, Switzerland}
\newcommand{\kassel}{Institut f\"{u}r Physik, Universit\"{a}t Kassel, Heinrich-Plett-Stra{\ss}e 40, 34132 Kassel, Germany}
\newcommand{\madrid}{Departamento de F\'{i}sica Te\'{o}rica de la Materia Condensada and Condensed Matter Physics Center (IFIMAC), Universidad Aut\'{o}noma de Madrid, E-28049 Madrid, Spain}
\newcommand{\eucor}{EUCOR Center for Quantum Science and Quantum Computing, Albert-Ludwigs-Universit\"{a}t Freiburg, Hermann-Herder-Stra{\ss}e 3, D-79104, Freiburg, Germany}

\author{Frieder Lindel}
\email{flindel@phys.ethz.ch}
\thanks{present address: Institute for Theoretical Physics, ETH Zürich, Zürich 8093, Switzerland.}
\affiliation{\madrid}
\affiliation{\freiburg}
\author{Stefan Yoshi Buhmann}
\affiliation{\kassel}
\author{Andreas Buchleitner}
\email{andreas.buchleitner@physik.uni-freiburg.de}
\affiliation{\freiburg}
\affiliation{\eucor}
\author{Edoardo G.\ Carnio}
\email{edoardo.carnio@physik.uni-freiburg.de}
\affiliation{\freiburg}
\affiliation{\eucor}

\date{\today}

\begin{abstract}
The rate of $m$-photon transitions in matter, induced by an incident light field, depends on the field's $m$th order coherence function. Consequently, the coherence properties of the light field may be shaped to increase the rate of multi-photon transitions. Here, we determine the optimal state of a weak fixed-intensity, narrow-band incident light field, with a restricted maximal photon number, that optimally drives $m$-photon transitions in the case of a short-lived atomic multilevel system. We show that, in this case, no quantum properties of the light field need to be exploited, but that classical mixtures of coherent states are optimal.
\end{abstract}

\maketitle

\section{Introduction}

The control of quantum systems lies at the heart of many quantum technologies. Standard control is achieved by using classical light fields, whose amplitude, polarization, frequencies, and relative phases are used as control knobs \cite{shapiro_quantum_2011}. Improving the efficiency of such control protocols is crucial to avoid, e.g., decoherence during a targeted unitary evolution, or other side effects such as heating or photodamage in the sample. 

To enhance the efficiencies of different control protocols, it has been suggested to harness the quantum properties of the control field \cite{mukamel_roadmap_2020}. A paradigmatic example is given by multi-photon absorption processes in the weak light--matter coupling regime. Here, only the lowest non-vanishing order of perturbation theory in the light--matter coupling has to be considered \cite{dorfman_nonlinear_2016}. It was found theoretically \cite{lambropoulos_coherence_1966,mollow_two-photon_1968,lambropoulos_field-correlation_1968,agarwal_field-correlation_1970} and experimentally \cite{jechow2013enhanced,spasibko_multiphoton_2017} that, in this regime, light with strong number fluctuations, such as squeezed or thermal states of light, can drive multi-photon absorption processes in free space more efficiently than quasiclassical coherent states with the same intensity. If multimode control fields are employed, two frequency-entangled photons can induce two-photon transitions more efficiently than any classically correlated photon pair \cite{georgiades_nonclassical_1995,dayan_two_2004,lee_entangled_2006,schlawin_theory_2017}. The implications of these enhanced control protocols have been discussed for a range of different applications, such as spectroscopy \cite{polzik_spectroscopy_1992,szoke_entangled_2020,mukamel_roadmap_2020}, quantum imaging \cite{gilaberte2019perspectives}, or quantum sensing \cite{carreno_exciting_2015,munoz2021quantum}. More recently, the potential of quantum states of light in case the control field is strongly coupled to the target system, as happens, e.g., in a cavity, \cite{castro_optimal_2019,lindel2023quantized} or in strongly driven systems \cite{gorlach2023high,even2023photon}, have been analyzed.

In our present contribution, we consider the following question: Which is the \textit{optimal} initial state of a single-mode field to drive an $m$-photon absorption process in the weak (perturbative) light--matter coupling regime? By fixing the average number of photons in the (quantized) field, and optimizing over a finite-dimensional part of its Hilbert space, we find that the optimal initial field states can be very well approximated by classical mixtures of coherent states. This implies that no quantum feature is needed to induce maximal enhancement of single-mode-driven $m$-photon absorption rates in the perturbative regime, but classical mixtures of quasi-classical coherent states are sufficient. The origin of the enhancement is traced back to the nonlinear scaling of the $m$-photon transition rate with the number of photons in the light field.

\section{Multi-photon absorption processes}

We consider a single atom with dipole operator $\hat{d}$ that is driven by a narrowband field $\hat{E}(t)$ with central frequency $\omega_0$ and bandwidth $\Delta \omega \ll \omega_0$. The light--matter interaction Hamiltonian is given by
\begin{align}
    \hat{H}_I = \hat{d}(t) [\hat{E}(t)+\hat{E}^\dagger(t)] ,
\end{align}
where $\hat{E}(t) $ denotes the positive frequency part of the field propagating in a single direction with cross section $A$, which is given in terms of creation and annihilation operators $\hat{a}^{(\dagger)}(\omega)$ via \cite{carnio_how_2021}
\begin{align}
    \hat{E}(t) =  \mi \int_0^\infty   \dif\omega \left(\frac{\hbar \omega}{4\pi c \epsilon_0 A} \right)^{1/2}\hat{a}(\omega) \me^{-\mi \omega t}.
\end{align}
We assume that the atom is initially in the ground state $\ket{G}$, with energy $\omega_G$, and there is a single atomic level $\ket{F}$, with energy $\omega_F$, which is resonant with an $m$-photon transition such that $\omega_F = \omega_G + m \omega_0$. While other discrete atomic levels may be present, those do not matter for the discussion that follows \cite{mollow_two-photon_1968}, and we therefore do not specify them further. For $m = 2$, one finds that the probability $p_F (t)$ of finding the system in the state $\ket{F}$ at times $t \gg (\Delta \omega)^{-1}$ (i.e., after the driving field has faded out), is given, in lowest-order perturbation theory in $\hat{H}_I$, by \cite{mollow_two-photon_1968} 
\begin{align} \label{eqP3Cont:TwoPhotAbsMultimode}
    p_F (t)= 2 t g^\prime \int \dif t^\prime \me^{2\mi \omega_F t^\prime - \gamma_F |t^\prime| } G^{(2)}(-t^\prime,-t^\prime,t^\prime,t^\prime).
\end{align}
Here, $g^\prime$ is a coupling constant (which encodes the electronic structure of the atom), and we have defined $G^{(2)}(-t,-t,t,t) = \braket{\hat{E}^\dagger(-t)\hat{E}^\dagger(-t)\hat{E}(t)\hat{E}(t)}$. Moreover, $\braket{\dots}$ indicates the expectation value with respect to the state of the field.
If the lifetime of the target state $\ket{F}$ is much shorter than the coherence time of the field---or equivalently, if the linewidth $\gamma_F$ of level $\ket{F}$ greatly exceeds the field bandwidth, $\gamma_F \gg \Delta \omega$~\cite{mollow_two-photon_1968}---the atomic system in state $\ket{F}$ decays on timescales much shorter than the pulse coherence time. In this regime, $p_F$ therefore only depends on $ G^{(2)}(0,0,0,0) \equiv G^{(2)}$, and Eq.~\eqref{eqP3Cont:TwoPhotAbsMultimode} simplifies to~\cite{mollow_two-photon_1968,agarwal_field-correlation_1970}
\begin{align} \label{eq:pftoG2}
p_F(t) \propto G^{(2)} t.
\end{align} 
Here, $G^{(2)}$ is the equal-time, normally-ordered, unnormalized second-order coherence function of the field,
\begin{align} \label{eq:G2Def}
G^{(2)}=\langle \big(\hat{a}^{\dagger }\big)^2 \hat{a}^2 \rangle,
\end{align}
where $\hat{a}$ is the annihilation operator of the field mode with frequency $\omega_0$. Similarly, one finds that the probability for an $m$-photon transition is proportional to the $m$th order coherence function $G^{(m)}$ \cite{agarwal_dynamical_nodate,decuing1964,lambropoulos_coherence_1966,lecompte1975}, defined by 
\begin{align} \label{eq:GmDef}
G^{(m)}=\langle \big(\hat{a}^{\dagger }\big)^m \hat{a}^m \rangle.
\end{align}

All following results in this contribution are based on Eq.~\eqref{eq:pftoG2}, and thus apply whenever this equation is valid. As we have just discussed, this is the case for a short-lived atomic multilevel system that is driven by a narrow-band field. 

\section{Optimal control of multi-photon absorption processes}

To find the optimal field state that maximizes the $m$-photon transition probability, one has to maximize the normally-ordered, $m$th-order coherence function $G^{(m)}$ of the field over all possible single-mode quantum field states. However, $G^{(m)}$ is, in general, not bounded from above. For example, for coherent states of light, one has $G^{(m)} \propto (n_\mathrm{av})^m$ with the average number of photons $n_\mathrm{av}=\langle \hat{a}^\dagger \hat{a} \rangle$. To find the optimally tailored quantum statistics that drive the $m$-photon transition with the highest efficiency, we therefore restrict our discussion to field states with fixed $n_\mathrm{av}$. This constraint still does not bound $G^{(m)}$ from above in general. As shown in Ref.~\cite{zubizarreta2017structure}, such a bound exists if we restrict the infinite-dimensional Hilbert space of the single-mode field to a sub-space $\mathcal{H}_{N_\mathrm{max}}$ spanned by all Fock states with at most a finite maximal number of photons $N_\mathrm{max}$. The maximal value of $G^{(2)} $ and $G^{(m)}$ is then given by \cite{zubizarreta2017structure}
\begin{align}
G^{(2)}  & \le n_\mathrm{av} (N_\mathrm{max} -1), \\
G^{(m)}  & \le \frac{(N_\mathrm{max} -2 )!}{(N_\mathrm{max} -m )!} G^{(2)},
\end{align}
and is attained by so-called coin states \cite{zubizarreta2017structure}
\begin{multline} \label{eqP3Cont:coinState}
\ket{\mathrm{coin}}  = \frac{1}{\sqrt{N_\mathrm{max}}} \left( \sqrt{N_\mathrm{max} - n_\mathrm{av}}\ket{0}  + \me^{\mi \phi} \sqrt{n_\mathrm{av}}\ket{N_\mathrm{max}} \right).
\end{multline}
Note that the value of $G^{(2)}$ is independent of the relative phase $\phi$ between the states $\ket{N_\mathrm{max}}$ and $\ket{0}$ in Eq.~\eqref{eqP3Cont:coinState}. These coin states are the states with the broadest photon number distribution around the average value $n_\mathrm{av}$, which still belong to the Hilbert space with maximally $N_\mathrm{max}$ photons. Consequently, they exhibit the largest possible fluctuations of the average photon number, intuitively explaining their maximization of $G^{(2)}$, which is known to quantify photon number fluctuations \cite{loudon2000quantum}. 

Which is the increase in efficiency achieved when using coin states instead of coherent states $\ket{\alpha}$ with average photon number $|\alpha|^2 = n_\mathrm{av}$? To answer this question, we consider the ratio 
\begin{align} \label{eqP3Cont:AdvantageCoin}
\frac{G^{(m)}_\mathrm{coin}}{G^{(m)}_\mathrm{coh}}  & = \frac{(N_\mathrm{max} -1 )!}{(N_\mathrm{max} -m )! n_\mathrm{av}^{m-1}} \approx \left( \frac{N_\mathrm{max}}{n_\mathrm{av}} \right)^{m-1},
\end{align}
where we defined $G^{(m)}_\mathrm{coh} =\bra{\alpha} (\hat{a}^\dagger)^m \hat{a}^m \ket{\alpha}$, $G^{(m)}_\mathrm{coin} =\bra{\mathrm{coin}} (\hat{a}^\dagger)^m \hat{a}^m \ket{\mathrm{coin}}$, and assumed $N_\mathrm{max} \gg m$ to obtain the approximate result.

We thus find that coin states are the optimal states to drive $m$-photon absorption processes in the single-mode weak-coupling limit, with enhancement factor $ (N_\mathrm{max}/n_\mathrm{av})^{m-1}$ of the absorption rate, in comparison to coherent states with the same average number of photons.
Generating coin states requires the preparation of a Fock state with a large photon number, which is a formidable experimental challenge. As we show in the following, however, the coin state can be very well approximated by a classical mixture of coherent states if $N_\mathrm{max} \gg 1$.

First, we notice that, by its very definition \eqref{eq:GmDef}, $G^{(m)}$ does not depend on the coherences (off-diagonal elements of the density matrix) of the state of the single-mode field in the Fock basis. Consequently, the same value of $G^{(m)}$, and thus the same efficiency in driving $m$-photon absorption processes, is obtained if, instead of the coherent superposition as given by by the coin state \eqref{eqP3Cont:coinState}, one uses the classical mixture
\begin{multline} \label{eqP3Cont:coinDens}
\hat{\rho}_\mathrm{coin} = \frac{1}{N_\mathrm{max}} \left[(N_\mathrm{max} - n_\mathrm{av})\ket{0}\bra{0} \right. \\ \left. 
+ n_\mathrm{av} \ket{N_\mathrm{max}}\bra{N_\mathrm{max}}\right].
\end{multline}
Next, we further approximate $\ket{N_\mathrm{max}}$ in the coin state $\hat{\rho}_\mathrm{coin}$ by coherent states $\ket{\alpha_{N_\mathrm{max}} }$ with $|\alpha_{N_\mathrm{max}}|^2 = N_\mathrm{max}$ (note that, formally, $ \ket{\alpha_{N_\mathrm{max}}} \notin \mathcal{H}_{N_\mathrm{max}} $\footnote{Strictly speaking, the coherent state $\ket{\alpha_{N_\mathrm{max}}}$ does not belong to the Hilbert space with maximally $N_\mathrm{max}$ photons, since the upper tail of its photon-number distribution will extend beyond $N_\mathrm{max}$. The state \eqref{eqP3Cont:coinDensCoh} is therefore not the optimal state found by optimizing $G^{(m)}$ over the finite-dimensional Hilbert space $\mathcal{H}_{N_\mathrm{max}}$.}):
\begin{multline} \label{eqP3Cont:coinDensCoh}
\hat{\rho}_\mathrm{coin,coh} = \frac{1}{N_\mathrm{max}} \left[(N_\mathrm{max} - n_\mathrm{av})\ket{0}\bra{0}\right. \\ \left.  + n_\mathrm{av} \ket{\alpha_{N_\mathrm{max}}}\bra{\alpha_{N_\mathrm{max}}}\right].
\end{multline}
Here, we have assumed $\braket{\alpha_{N_\mathrm{max}}| 0} \approx 0$, which is valid for $N_\mathrm{max} \gg 1$. For the state $\hat{\rho}_\mathrm{coin,coh}$, we find that the $m$-point coherence function (normalized by $G^{(m)}_\mathrm{coh}$) reads
\begin{align}  \label{eqP3Cont:AdvantageCoinCoh}
\frac{G^{(m)}_\mathrm{coin,coh}}{G^{(m)}_\mathrm{coh}}  & =\left( \frac{N_\mathrm{max}}{n_\mathrm{av}} \right)^{m-1},
\end{align}
in agreement with Eq.~\eqref{eqP3Cont:AdvantageCoin}, in the limit $N_\mathrm{max} \gg m$.

\begin{figure}
\includegraphics[width=\columnwidth]{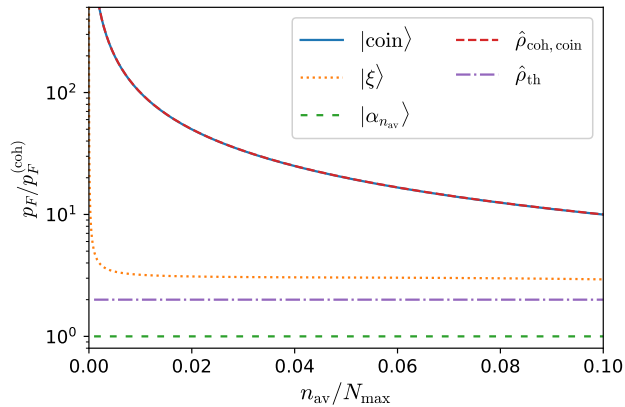}
\caption{\textit{Two-photon absorption rates.} Probability $p_F$ of a two-photon transition, as a function of the average number $n_\mathrm{av}$ of photons initially in the field. We consider different initial states of the light field: the coin state $\ket{\mathrm{coin}}$, Eq.~\eqref{eqP3Cont:coinState}; the coin state approximated by a mixture of coherent states $\hat{\rho}_\mathrm{coin,coh}$, Eq.~\eqref{eqP3Cont:coinDensCoh}; a squeezed vacuum state $\ket{\xi} = \me^{(\xi^\ast \hat{a}^2 + \xi (\hat{a}^\dagger)^2)/2} \ket{0}$ determined by the squeezing parameter $\xi$; a thermal state $\hat{\rho}_\mathrm{th} = \me^{-\hbar \omega a^\dagger a/k_B T } / \mathrm{tr}\{\me^{-\hbar \omega a^\dagger a/k_B T } \} $ with Boltzmann constant $k_B$ and temperature $T$, for which $G^{(2)}/G^{(2)}_\mathrm{coh} = 2, \forall T$; a coherent state $\ket{\alpha_{n_\mathrm{av}}}$ with $|\alpha_{n_\mathrm{av}}|^2 = n_\mathrm{av}$. For $\ket{\xi}$, we obtained $G^{(2)}$ numerically for $\xi \in [0,2.65]$, which leads to the values of $n_\mathrm{av} = \mathrm{sinh}^2(|\xi|)$ shown in the Figure. If $\xi \gg 1$, we find $G^{(m)}/G^{(m)}_\mathrm{coh} \approx (2m-1)!!$ \cite{janszky_many-photon_1987}, i.e., $G^{(2)}/G^{(2)}_\mathrm{coh} \approx 3 $. We restrict $n_\mathrm{av}$ to a regime in which all considered field states still lie within the restricted Hilbert space $\mathcal{H}_{N_\mathrm{max}}$, with $N_\mathrm{max} = 500$ (they have therefore negligible overlap with Fock states $\ket{n}$ with $n>N_\mathrm{max}$).}
\label{fig:SingleMode}
\end{figure}

We have thus shown that a classical mixture of the vacuum with a coherent state optimally drives $m$-photon absorption processes in the perturbative regime. As illustrated in Fig.~\ref{fig:SingleMode}, this state is better suited than coherent, squeezed or thermal states with the same average number of photons (i.e., with the same intensity).

\section{Discussion and conclusion}

What is the origin of the advantage of the state $\hat{\rho}_\mathrm{coin,coh}$ (compared to just a coherent state) for driving $m$-photon transitions? As a statistical mixture, the state $\hat{\rho}_\mathrm{coin,coh}$ in Eq.~\eqref{eqP3Cont:coinDensCoh} formally describes the situation in which one averages over many experiments. In each experiment, there is a probability $n_\mathrm{av}/N_\mathrm{max}$ of employing a coherent laser driving with an average photon count of $N_\mathrm{max} > n_\mathrm{av}$, while in the remaining cases, the driving field is blocked, exposing the sample solely to the field's vacuum state. The $m$-photon excitation probability $p_m $ is thus proportional to $ (n_\mathrm{av}/N_\mathrm{max})G^{(m)}$, where $G^{(m)}$ is the coherence function of the driving field with on average $N_\mathrm{max}$ photons, given by $G^{(m)} = N_\mathrm{max}^m$. Consequently, $p_m \propto (n_\mathrm{av}/N_\mathrm{max})G^{(m)} = n_\mathrm{av} N_\mathrm{max}^{m-1}$, whereas for a coherent state with $n_\mathrm{av}$ photons, $p_m \propto n_\mathrm{av}^m$.

This demonstrates that the enhancement of $m$-photon absorption processes via enhanced number fluctuations arises from the fact that the probability of an $m$-photon absorption process, with $m \ge 2$, scales faster than linearly with the number of photons in the state. For a fixed average photon count $n_\mathrm{av}$, it is thus advantageous to maintain a low, yet non-zero, probability of encountering a state with a significant number of photons $n\gg n_\mathrm{av}$, compared to having a steady field state with $n_\mathrm{av}$ photons. This phenomenon is purely classical. We emphasize that our results do not apply to multi-mode driving of $m$-photon transitions, e.g., via two-photon states entangled in the frequency domain \cite{georgiades_nonclassical_1995,dayan_two_2004,lee_entangled_2006,schlawin_theory_2017}.

In conclusion, we have shown that quantum features provide no advantage for optimally driving $m$-photon transitions between two discrete electronic states in the single-mode and weak-coupling regime considered here. Instead, a classical mixture of the vacuum with a coherent state, which approximates coin states \eqref{eqP3Cont:coinState} very well, is found to be the optimal field state.

\begin{acknowledgments}
The authors are grateful to Frank Schlawin and Peter Labropoulos for fruitful discussions and constructive comments. F.L. acknowledges support from the Studienstiftung des deutschen Volkes. This work was supported by the Spanish Ministry for Science and Innovation-Agencia Estatal de Investigación (AEI) through the grant EUR2023-143478. We also acknowledge financial support from the European Union’s Horizon Europe Research and Innovation Programme through agreement 101098813(SCOLED).
\end{acknowledgments}

%

\end{document}